\documentclass[galaxies,article,accept,moreauthors,pdftex]{Definitions/mdpi} 
\usepackage{amsmath}
\firstpage{1} 
\pubvolume{9}
\issuenum{4}
\articlenumber{95}
\pubyear{2021}
\copyrightyear{2021}
\externaleditor{Academic Editors: Elena Moretti \mbox{and Francesco Longo}} % To Editors: Please add.
\datereceived{30 September 2021} 
\dateaccepted{4 November 2021} 
\datepublished{7 November 2021} 
\hreflink{https://doi.org/\linebreak 10.3390/galaxies9040095}
\Title{Accounting for Selection Bias and Redshift Evolution in GRB Radio Afterglow Data}
\TitleCitation{Accounting for Selection Bias and Redshift Evolution in GRB Radio Afterglow Data}
\Author{{Maria Dainotti}~$^{1,2,3,}$*$^{,\dagger}$, 
Delina Levine~$^{4,\dagger}$, 
Nissim Fraija~$^{5}$ and 
Poonam Chandra~$^{6}$}
\AuthorNames{Maria Dainotti, Delina Levine, Nissim Fraija and Poonam Chandra}
\AuthorCitation{Dainotti, M.; Levine, D.; Fraija, N.; Chandra, P.}
\address{%
$^{1}$ \quad National Astronomical Observatory of Japan, 2-21-1 Osawa,  Tokyo 181-8588, Japan\\
$^{2}$ \quad Space Science Institute, {Boulder, CO 80301}, USA\\%Please add zip code.
$^{3}$ \quad Department of Astronomical Science, School of Physical Sciences, The Graduate University for Advanced Studies, SOKENDAI, Shonankokusaimura, Miura District, Hayama  240-0193, Japan \\
$^{4}$ \quad Department of Astronomy, University of Maryland, College Park, MD 20742, USA; dmlevine@umd.edu	\\
$^{5}$ \quad Instituto de Astronomia, {Universidad Nacional Aut\'{o}noma de M\'{e}xico}, Mexico City 04510, Mexico; nifraija@astro.unam.mx	\\
$^{6}$ \quad National Centre for Radio Astrophysics, Tata Institute of Fundamental Research, {Pune} 411007, %Please confirm whether it is a city 
India; poonam@ncra.tifr.res.in}

\corres{Correspondence: {maria.dainotti@nao.ac.jp}}%Please confirm who is the correspondent author and mark as asterisk.
\firstnote{These authors contributed equally to this work.}

\abstract{Gamma-ray Bursts (GRBs) are highly energetic events that can be observed at extremely high redshift. However, inherent bias in GRB data due to selection effects and redshift evolution can significantly skew any subsequent analysis. We correct for important variables related to the GRB emission, such as the burst duration, $T_{90}^*$, the prompt isotropic energy, $E_{\rm iso}$, the rest-frame end time of the plateau emission, $T_{\rm a, radio}^*$, and its correspondent luminosity $L_{\rm a, radio}$, for radio afterglow. In particular, we use the Efron–Petrosian method {presented in 1992} %MDPI: Please do not cite reference in abstract.
for the correction of our variables of interest. Specifically, we correct $E_{\rm iso}$ and $T_{90}^*$ for 80 GRBs, and $L_{\rm a, radio}$ and $T_{\rm a, radio}^*$ for a subsample of 18 GRBs that present a plateau-like flattening in their light curve. Upon application of this method, we find strong evolution with redshift in most variables, particularly in $L_{\rm a, radio}$, with values similar to those found in past and current literature in radio, X-ray and optical wavelengths, indicating that these variables are susceptible to observational bias. This analysis emphasizes the necessity of correcting observational data for evolutionary effects to obtain the intrinsic behavior of correlations to use them as discriminators among the most plausible theoretical models and as reliable \mbox{cosmological tools.}}


\begin{document}
\section{Introduction}\label{intro}

Gamma-ray bursts (GRBs) can be observed at extremely high redshift, making them a valuable tool for studies of the early Universe. The ability to observe these highly energetic objects at {redshifts out to $z = 9.4$}~\citep{2011ApJ...736....7C, 2009MNRAS.400..775C,2009Natur.461.1258S, 2009Natur.461.1254T, 2010MNRAS.408.1181C, 2014ApJ...783..126P, 2013ApJ...774..157D, 2013MNRAS.436.3640L, 2014ApJ...781....1L, 2016ApJ...825..135M, 2018ApJ...865..107T} has created significant interest in using them as standardizable candles, similar to Type Ia supernovae. However, observations of GRBs have shown a very diverse population with few common characteristics. %MDPI： Please confirm the citation of ref 1 and update the order.

Phenomenologically, GRBs are characterized by the main event, called the prompt emission, which is usually observed in gamma-rays, hard X-rays and sometimes in optical, while the afterglow is the counterpart in soft X-rays ($\approx$66\% of observed GRBs), in optical ($\approx$38\% of observed GRBs) and sometimes in radio ($\approx$6.6\% of observed GRBs). GRB radio afterglows are very difficult to observe, indeed, similar to the X-ray observations which are characterized by the detector limits, and additional difficulties rise due to the limited allocated time for the follow-up observations in the radio band after the GRB trigger. Bursts are classified following the duration of the prompt episode ($T_{90}$). The population of short GRBs (sGRBs) usually has harder spectra and a duration of less than 2 s. In contrast, the population of long GRBs (lGRBs) has softer spectra and a duration larger than 2 s~\citep{2013FrPhy...8..661G}. However, this classification is still in debate, and so in some bursts, it is not clear if GRBs with intermediate features belong to short or long GRBs~\citep{2006ApJ...643..266N, 2011ApJ...735...23N}. LGRBs are associated with the core collapses of dying massive stars~\citep{1993ApJ...405..273W, 1998ApJ...494L..45P} and sGRBs by merging two compact objects; a black hole (BH) with a neutron star (NS) and two NSs~\citep{1992Natur.357..472U, 1992ApJ...392L...9D, 2011MNRAS.413.2031M}.

\textls[-5]{A crucial breakthrough in the analysis of GRB features is the discovery of the plateau emission by the Neils Gehrels Swift Observatory~\citep{2004ApJ...611.1005G}. The plateau emission is a flat part of the lightcurves which follows the decay phase of the prompt emission~\citep{2007ApJ...669.1115S, 2006ApJ...647.1213O, 2007ApJ...662.1093W, 2015AdAst2015E..22P}. In the current paper we focus on properties resulting from this plateau emission, as well as both the prompt and afterglow emission. In general, attempts have been made to find standardizable properties, such as a plateau of GRBs or through prompt and afterglow correlation studies. We here mention a few of them:~\citet{2004ApJ...609..935Y}, \mbox{\citet{2007AdSpR..40.1244D}}, \mbox{\citet{2007RSPTA.365.1385G}},~\citet{2008MNRAS.391L..79D},~Amati et al. \cite{2008MNRAS.391..577A}, Dainotti et al. \cite{2010ApJ...722L.215D, 2011ApJ...730..135D, 2011MNRAS.418.2202D, 2016ApJ...825L..20D, 2017A&A...600A..98D}, \mbox{\citet{2019gbcc.book.....D}}.}

However, it is clear from past and current studies~\citep{2007ApJ...667.1024K, 2011MNRAS.411.1843S, 2015MNRAS.451..126S, 2015MNRAS.449L...6C} that observations of GRBs are further susceptible to selection bias and evolutionary effects, which may change the results of any subsequent analysis and can substantially impact the results related to cosmological application of GRB relations~\citep{2013MNRAS.436...82D}. In GRB studies, it is, therefore, crucial to know whether the studied correlations are intrinsic or artificially created as a result of observational biases and redshift evolution. ``Redshift evolution" is the dependence of the variable of interest on redshift, and thus ``independent of redshift" indicates the absence of such evolution. 

In the study of GRB correlations, all variables must be computed in the rest-frame, as we are comparing objects at different epochs. This introduces another source of redshift dependence included in the definition of luminosity distance:
\begin{equation}
D_L=(1+z) \frac{c}{H_0} \int_{0}^{z} \frac{dz^{'}}{\sqrt{\Omega_M (1+z')^3 + \Omega_\Lambda}}\,,
\label{dl}
\end{equation}
where $H_0$ is the Hubble constant at the present day and $\Omega_M$ is the matter density in a flat Universe assuming the equation of state parameter w = $-$1. Indeed, usually one of the variables in the correlation is either a luminosity or energy which, by definition, depends on the luminosity distance. Ideally, all correlations we use must be corrected for redshift evolution, if any, requiring the removal of any existing redshift dependence.

There do exist statistical techniques that are capable of correcting for these effects, as well as correcting for data truncation from detector limits~\citep{1971MNRAS.155...95L, 1992ApJ...399..345E, 1998astro.ph..8334E}. Among the methods to remove evolution, we consider here the Efron–Petrosian (EP)~\citep{1992ApJ...399..345E} method. The EP method is a well-established example of these kinds of techniques, and has been used to recover intrinsic relationships in many correlations in the past~\citep{2000ApJ...534..227L, 2011ApJ...743..104S, 2013ApJ...774..157D, 2013MNRAS.436...82D, 2019MNRAS.488.5823L,2020MNRAS.494.4371L, 2021ApJ...914L..40D}. 

\citet{2000ApJ...534..227L} discuss the correlation between $E_p$, or the peak of the $\nu F_\nu$ spectrum, with flux and fluence in GRBs, later investigated in the rest-frame and known as the $E_p-E_{\text{iso}}$ relation~\citep{2002A&A...390...81A}. A further modification of this relation is the one discovered by~\citet{2004ApJ...609..935Y} in which $E_p$ is correlated with the prompt isotropic luminosity, $L_{\text{iso}}$. The EP method provides an explanation on how to perform analysis on truncated data, and in~\citet{2002A&A...390...81A, 2004ApJ...609..935Y} it is illustrated that the method is capable of recovering the correlation present in the original ``parent" sample with the truncated data. 

This technique has been further explored regarding the luminosity function and formation rate of sGRBs in a recent study by~\citet{2021ApJ...914L..40D}. They look at the intrinsic distributions of these variables using the EP method, and introduce a method of accounting for incompleteness of redshift data with the Kolmogorov–Smirnov (KS) test (this is described in more detail in Section \ref{limits}). They find a strong evolution of luminosity with redshift, emphasizing the necessity of this correction. 
The analysis presented in \mbox{\citet{2021ApJ...914L..40D}} is also relevant, as it emphasizes that both sGRBs and lGRBs undergo strong redshift~evolution.

It should also be noted that though this method is mainly applied to GRB correlation studies, it has been also successfully applied in studies of Active Galactic Nuclei as well~\citep{2011ApJ...743..104S}.

Among GRB correlations in particular we focus our attention to the rest-frame time at the end of the plateau emission, $T^{*}_{\rm a, radio}$, and its correspondent luminosity $L_{\rm a, radio}$, this is an extension in radio of the so-called 2D Dainotti relation in X-rays~\citep{2008MNRAS.391..577A, 2010ApJ...722L.215D, 2011ApJ...730..135D} and optical~\citep{2020ApJ...905L..26D}. For the very recent analysis on the 2D Dainotti relation in radio see Levine et~al. (2021) in preparation. For a review of the subject of GRB correlations and selection biases in the prompt and afterglow see~\citet{2017NewAR..77...23D, 2018AdAst2018E...1D, 2018PASP..130e1001D, 2007AdSpR..40.1244D, 2019gbcc.book.....D, 2007RSPTA.365.1385G}.

One of the main problems in the application of GRB relationships as theoretical model discriminators and as cosmological tools is the fact that correlations must be intrinsic to the physics and not induced by biases.
There are several examples of how the correlations are used to interpret theoretical models both in the prompt and afterglow emission. The photospheric emission and the Comptonization models~\citep{2012ApJ...752..116T, 2017SSRv..207...87B, 2020ApJ...897..145D, 2011MNRAS.415.1663T, 2006ApJ...646L..25Z, 2019NatCo..10.1504I, 2015MNRAS.454L..31A, 2013MNRAS.432.3237G,  2014ApJ...789..159I} are the two main models used to test the $E_{peak}$--$E_{\text{iso}}$ and the~\citet{2004ApJ...609..935Y} correlations, the latter between $E_{peak}$ and the isotropic energy in the prompt emission.
Otherwise, the parameter space pinpointed by those correlations can be the effect of selection biases and not of the true underlying physics. To this end, it is necessary to apply these correlations as model discriminators only after correction for such biases. Indeed, for example the plateau emission in X-rays and optical, which reconciles with the existence of the 2D Dainotti relation, can be derived through the equations of a fast rotating NS, the so-called magnetar model~\citep{1992Natur.357..472U, 1992ApJ...392L...9D, 2011MNRAS.413.2031M, 2011A&A...526A.121D, 2013MNRAS.430.1061R, 2021ApJ...918...12F}. In~\citet{2014MNRAS.443.1779R, 2015ApJ...813...92R, 2018ApJ...869..155S} the derivation of the parameter space of the magnetic field and spin period have been computed accounting for selection biases and redshift evolution. The current status in the literature is that only
a few correlations have been corrected for selection biases and evolutionary effects through the EP method, such as ~\citet{2013ApJ...774..157D, 2015ApJ...800...31D, 2017ApJ...848...88D, 2020ApJ...904...97D, 2021ApJ...914L..40D}.

Specifically,~\citet{2013ApJ...774..157D} examine this correlation in X-ray for a sample of 101 GRBs that present a plateau, or flattening, in their light curves. After correction for evolutionary effects using the EP method, they conclude that the observed correlation is intrinsic at the 12 $\sigma$ level. In mimicking the evolution of each variable with redshift, they tested both a simple and more complex model, finding similar results in both cases.~\citet{2013MNRAS.436...82D} further examine the importance of these corrections when studying the cosmological properties of GRBs, applying the EP method to a simulated correlation between luminosity and time at the end of the plateau emission for 101 GRBs and testing whether a 5 $\sigma$ deviation from the intrinsic values strongly changes the cosmological results. They demonstrated that their results change with this deviation by 13\% regarding the values of $\Omega_M$, emphasizing the necessity of applying such corrections. The problem of evolution of the variables and their correction is not only important for GRB-cosmology studies, but also for more general cosmological studies. Indeed, in~\citet{2021ApJ...912..150D} it has been shown that there is indication of a possible evolution even of the Hubble constant. If this is not due to selection biases, a new physical cosmological model which relies on alternative theories, such as the modified theory of gravity, must be accounted for.

Regarding the correlations in GRB afterglows,~\citet{2019MNRAS.488.5823L, 2019ApJ...871..118L} discussed the correlation of not only luminosity with redshift, but also isotropic energy, $E_{\rm iso}$, $T_{90}^*$, and the jet opening angle, $\theta_j$, for a sample of 376 GRBs. They emphasize the difficulty of obtaining intrinsic values for these quantities due to inherent biases in observation methods, and additional truncation from detector limits that can introduce false correlations in the data~\citep{2021MNRAS.504.4192B}. They find strong evolution with redshift for each of these variables, indicating that achromatic properties of GRBs are also susceptible to selection bias. A further study by~\citet{2020MNRAS.494.4371L} discusses the evolution of $\theta_j$ with redshift in greater detail, using the EP method to recover the intrinsic behavior of the jet opening angle.

In this study, we seek to determine whether the strong evolution of $E_{\text{iso}}$ and $T_{90}^*$ vs. redshift initially found by~\citet{2019MNRAS.488.5823L, 2019ApJ...871..118L} is still the same for GRBs with observed radio afterglow. In addition to the isotropic energy, we apply the EP method to the luminosity, and break time in radio wavelengths to determine if these variables are strongly affected by inherent bias and evolutionary effects. 

We here point out that we are aware that the plateau sample is a subsample of a more extended population of plateaus that we cannot see. We have fixed the issue of the biases related to the redshift evolution and due to the selection threshold with the Efron–Petrosian method; however, we cannot account for the missing population of GRBs for which the follow-up has not been tackled. Nonetheless, it is crucial to discuss the $L_{\rm a,radio}$ versus the redshift, since this correlation has been studied in X-rays and optical extensively and it is important to investigate if this correlation holds true in radio with comparable slopes to optical and X-ray. The first step to investigate the correlation is to determine if the variables involved are subjected to redshift evolution and selection biases.

In summary, the main point of the paper is the study of the redshift evolution and the removal of selection biases through the EP method. The analysis of the true correlations can be done only if we first determine the evolution among the variables. The plateau emission has been extensively investigated in X-rays and optical, but so far there has not been a statistical analysis of the existence of the plateau in radio. The radio observations of the plateau emission can cast a light on whether or not the end time of the plateau is indeed a jet break. This point can be revealed only through such a study. The evaluation of the jet break allows one to better understand the evolution of the GRB and its physics in relation to the standard fireball model~\citep{1999PhR...314..575P, 1999ApJ...517L.109S} or other scenarios.

This paper is organized as follows: in Section \ref{methods}, we discuss the selection of our sample, as well as the formulation of the EP method and its application to our sample. In Section \ref{results}, we present the results of this analysis, and in Section \ref{discussion}, we discuss the implications of our study, as well as a comparison to previous studies, and present our conclusions.
 
%%%%%%%%%%%%%%%%%%%%%%%%%%%%%%%%%%%%%%%%%
\section{Methods}\label{methods}

\subsection{Variables of Interest}\label{variables}

The EP method uses a modified version of Kendall's $\tau$ statistics to test for independence of variables in a truncated data set. Here, $\tau$ is defined as: 
\begin{equation}
\tau =\frac{\sum_{i}{(\mathcal{R}_i-\mathcal{E}_i)}}{\sqrt{\sum_i{\mathcal{V}_i}}}
\label{tau}
\end{equation}
\textls[-15]{with  $R_i$ defined as the rank $\mathcal{E}_i=(1/2)(i+1)$ defined as the expectation value, and $\mathcal{V}_i=(1/12)(i^{2}+1)$ defined as the variance. For a more complete discussion of the method and the algebra involved, see~\citet{2011ApJ...743..104S, 2013ApJ...774..157D, 2017A&A...600A..98D, 2015MNRAS.451.3898D, 1992ApJ...399..345E, 2015ApJ...806...44P, 2002ApJ...565..182L}. Here, we use the EP method to determine the impact of redshift evolution and selection bias on four variables: $T_{90}^*$, where the star denotes the rest-frame, $E_{\rm iso}$, the radio light curve break time $T_{\rm a, radio}^*$, and the radio luminosity at the time of break $L_{\rm a, radio}$. These variables are considered to be they are pertinent to the correlations analyzed in Levine et~al. (2021 in preparation). Throughout our analysis, we consider these variables in logarithmic scale for convenience.}

We look at the $\log E_{\rm iso}$ and $\log T_{90}^*$ for a sample of 80 GRBs with observed radio afterglow published in the literature~\citep{2012ApJ...746..156C, 2013ApJ...767..161Z, 2015ApJ...814....1L, 2019ApJ...884..121L, 2015ApJ...812..122C, 2019MNRAS.486.2721B, 2014MNRAS.440.2059A, 2015ApJ...806...52S, 2018ApJ...859..134L, 2020ApJ...894...43K, 2016ApJ...833...88L, 2020ApJ...891L..15C, 2017ApJ...845..152B, 2017Sci...358.1579H, 2018Natur.554..207M, 2018ApJ...867...57R, 2018ApJ...856L..18M, 2021ApJ...907...60M, 2020MNRAS.496.3326R}. Values of $\log E_{\rm iso}$ are taken from the literature. If no $E_{\rm iso}$ value could be found, the $\log E_{\rm iso} $, in units of ergs, is calculated using the equation:
\begin{equation}
E_{\rm iso}=4\pi D_L^2 (z) S K\,,
\label{eqn:Eiso}
\end{equation}
where $S$ is the fluence in units of erg $\rm cm^{-2}$, $D_{\rm L}^2 (z)$ is the luminosity distance assuming a flat $\Lambda \textrm{CDM}$ model with $\Omega_M=0.3$ and $H_0=70$ $\mathrm{km}$ $\mathrm{s^{-1}}$ $\mathrm{\rm Mpc^{-1}}$ (see Equation~\eqref{dl}), and $K$ is the correction for cosmic expansion~\citep{Bloom_2001}:
\begin{equation}
K=\frac{1}{(1+z)^{1-\beta}}\,,
\label{eqn:Kcorr}
\end{equation}
with $\beta$ as the spectral index of the GRB. Fluence and $\beta$ values are taken from the literature. 
%\textcolor{magenta}{We do not consider two GRBs, GRB 050509C and 170105A, in our analysis of $\log E_{\rm iso}$ as the prompt emission was not observed.}

To analyze the impact of selection bias and redshift evolution for $\log L_{\rm a, radio}$ and $\log T_{\rm a, radio}^*$, we first fit each of the 80 GRBs with a broken power law (BPL) according to \mbox{the formulation:}
\begin{equation}
    F(t)  =
    \begin{cases}
    F_a (\frac{t}{T_a^*})^{-\alpha_1} & t < T_a^* \\
    F_a (\frac{t}{T_a^*})^{-\alpha_2} & t \geq T_a^*
    \end{cases}
\end{equation}
where $F_a$ refers to the flux at the break, $T_a^*$ refers to the rest-frame time of break, $\alpha_1$ refers to the slope before the break, and $\alpha_2$ refers to the slope after the break.
We can only obtain values of $\log L_{\rm a, radio}$ and $\log T_{\rm a, radio}^*$ for light curves that show a ``plateau'' feature, or a flattening of the light curve before a clear break. In our analysis, we consider a light curve to display a plateau if $|\alpha_1| < 0.5$. Therefore, we discard fits to light curves with scattered observations, unreliable error bars, or shapes incompatible with a BPL and plateau. In our subsequent analysis we include those light curves whose $\Delta \chi^2$ analysis of the BPL best-fit parameters are suitable following the~\citet{1978A&A....66..307A} methodology. After the rejection process, we are left with 18 GRBs that present a plateau and clear break in the light curve. 

\begin{figure}[H]
 %   \centering
    \begin{tabular}{cc}
        \includegraphics[width = 0.35\textwidth]{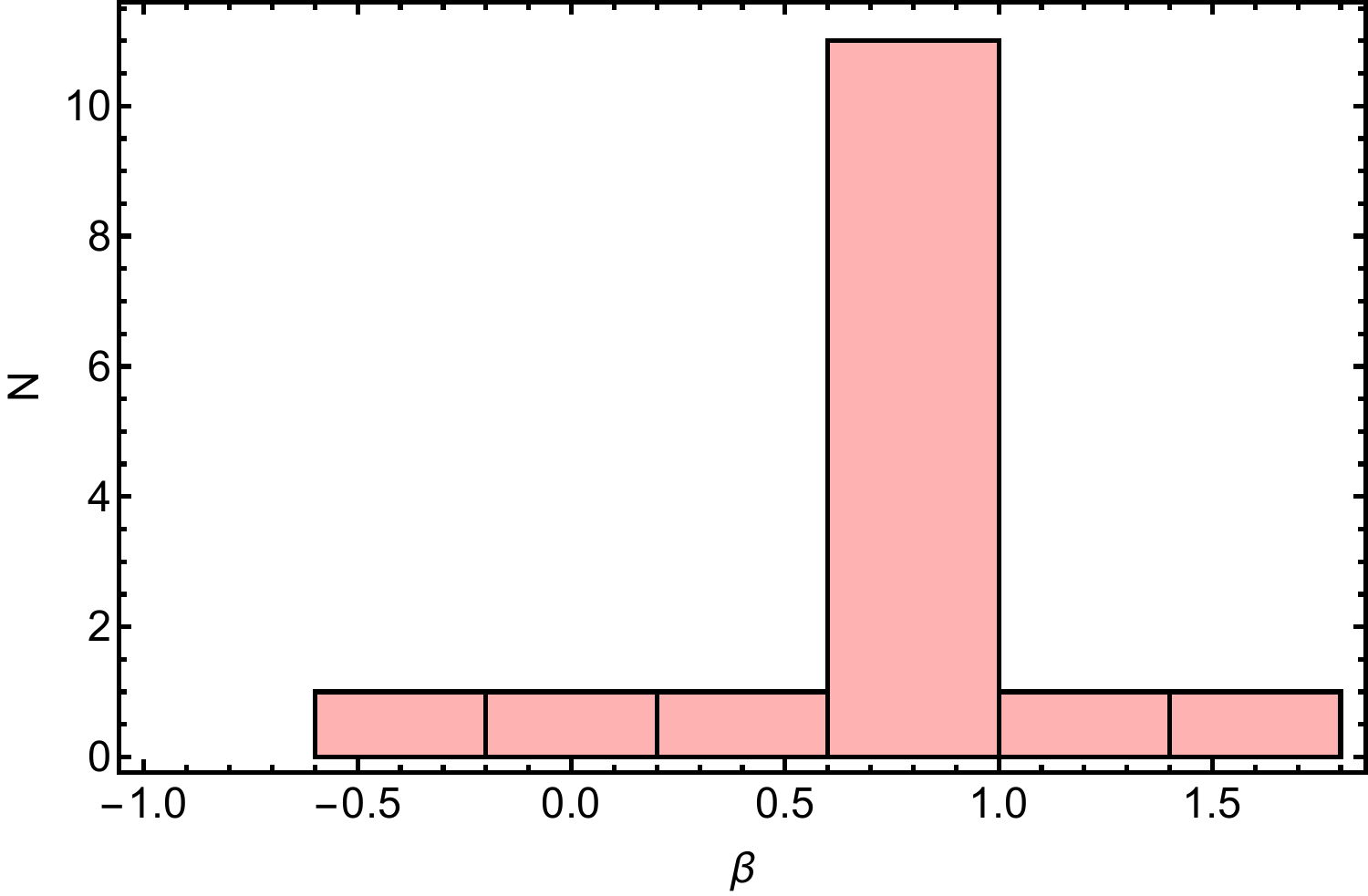}
    \end{tabular}
    \caption{Distribution of spectral indices ($\beta$) for sample of 18 GRBs that display a plateau in their light curve.}
    \label{fig:betahist}
\end{figure}

The luminosity $\log L_{\rm a, radio}$ in units of erg $s^{-1}$ is computed at time $\log T_{\rm a, radio}^*$ using \mbox{the equation:}
\begin{equation}
L_{a}=4\pi D_L^2 (z) F_{a} (T_a)K\,,
\label{eqn:LumTim}
\end{equation}
where $F_{a}$ is the observed flux at $T_{\rm a, radio}$, $D_{\rm L}^2 (z)$ is defined as in Equation~\eqref{eqn:Eiso}, and K is the k-correction: 
\begin{equation}
K=\frac{1}{(1+z)^{\alpha_1-\beta}}\,,
\label{eqn:kcorr}
\end{equation}
with $\beta$ as the radio spectral index and $\alpha_1$ as the fitted BPL temporal index before the break. $\beta$ values are taken from~\citet{2012ApJ...746..156C} or other literature---if no $\beta$ value could be found, the average of published spectral indices, $\beta = 0.902 \pm 0.17$, was assigned. We show the distribution of spectral indices for the plateau sample in Figure \ref{fig:betahist}.

\subsection{Limiting Fluxes, Fluences, and Times}\label{limits}

The EP method can overcome selection bias for a particular variable of interest, but it must first be determined if the variable is dependent or independent of redshift. 

\textls[-15]{It is then necessary to define limiting values for each of the variables. In \mbox{\citet{2013ApJ...774..157D}}, it was demonstrated that a good choice of limiting times and luminosities retains at least 90\% of the total sample. For time variables, $\log T_{90}^*$ and $\log T_{\rm a, radio}^*$ (in units of seconds), we define a general form for the limiting values as $\frac{T_{\rm min}}{(1+z)}$, where $T_{\rm min}$ is the minimum observed time. We need to choose a compromise between a limit which is representative of the population of data points, but still retains most of the sample size. It has been shown in Monte Carlo simulations in~\citep{2013ApJ...774..157D} that such a strategy with limiting values is accurate. For $\log T_{90}^*$, we find the best limiting duration to be $\log T_{{90_{\rm min, obs}}}^* = -0.54$ s, with a limiting boundary defined as $\frac{-0.30}{(1+z)}$ s, which excludes 5/80 ($<$$10\%$) GRBs. 
The limiting line for $\log T_{90}^*$ is shifted at a higher value to be allow the sample data to be representative of the whole population.
For $\log T_{\rm a, radio}^*$, we find the limiting time to be the observed minimum $\log T_{{{\rm a, radio}_{\rm min, obs}}}^* = 4.94$ s, thus defining the boundary as $\log T_{\rm a, radio}^* = \frac{4.94}{(1+z)}$ s, which does not exclude any data points.}

For the isotropic energy, we instead define the limiting energy according to the methodology of~\citet{2013ApJ...774..157D}, in which the limiting fluence should be representative of the population while including at least $90\%$ of the sample. We use the following formula: 
\begin{equation}
   E_{\rm iso, lim} = 4\pi D_L^2 (z) S_{\rm lim}\,, 
\end{equation}
where $S_{\rm lim}$ is the fluence limit. For our sample, we define $S_{\rm lim}$ as 6.3 $\times$ $10^{-8}$ $\mathrm{erg}$ $\mathrm{cm}^{-2}$. Applying this limit excludes 8/80 GRBs, which is 10\% of our sample. 
In all the method described here we use GRBs that have $\log E_{\rm iso} > \log E_{\rm iso,lim}$, $\log T^{*}_{90} > \log T^{*}_{90}$, ${\log T_{\rm a, radio}} > {\log T_{\rm a, radio, lim}}$, and ${\log L_{\rm a, radio}} > {\log L_{\rm a, radio, lim}}$.
For the luminosity, however, a caveat should be posed when we consider the total distribution of the parent population of GRBs with and without redshift (see~\citep{2021ApJ...914L..40D}).
 
Using the method presented in~\citet{2021ApJ...914L..40D}, we compare the parent sample of all GRBs with observed radio afterglow and known peak flux to a smaller ``subsample" of GRBs with known peak flux and known redshift. We then apply cuts to the data by defining limiting fluxes at regular intervals. Considering only the data with values above the limiting fluxes, fluences and time, we conduct a two-sample Kolmogorov–Smirnov (KS) test between the data of the total sample and the data for which the limiting cuts have been applied to determine the distribution of the probability that the subsample was drawn from the parent sample, as well as the geometric distance between the two samples as determined by the KS test. We take the limiting flux to be the value of $f_{\rm lim}$ where the probability as a function of limiting flux reaches a plateau in which the probability that two samples are drawn by the same population is 100\%. In our sample, we find this limit to be $\log f_{\rm lim} = -17.2$. We define the flux throughout our analysis in units of $\mathrm{erg}$ $\mathrm{cm}^{-2}$ $\mathrm{s}$.

\begin{figure}[H]
    \begin{tabular}{cc}
        \includegraphics[width = 0.35\textwidth]{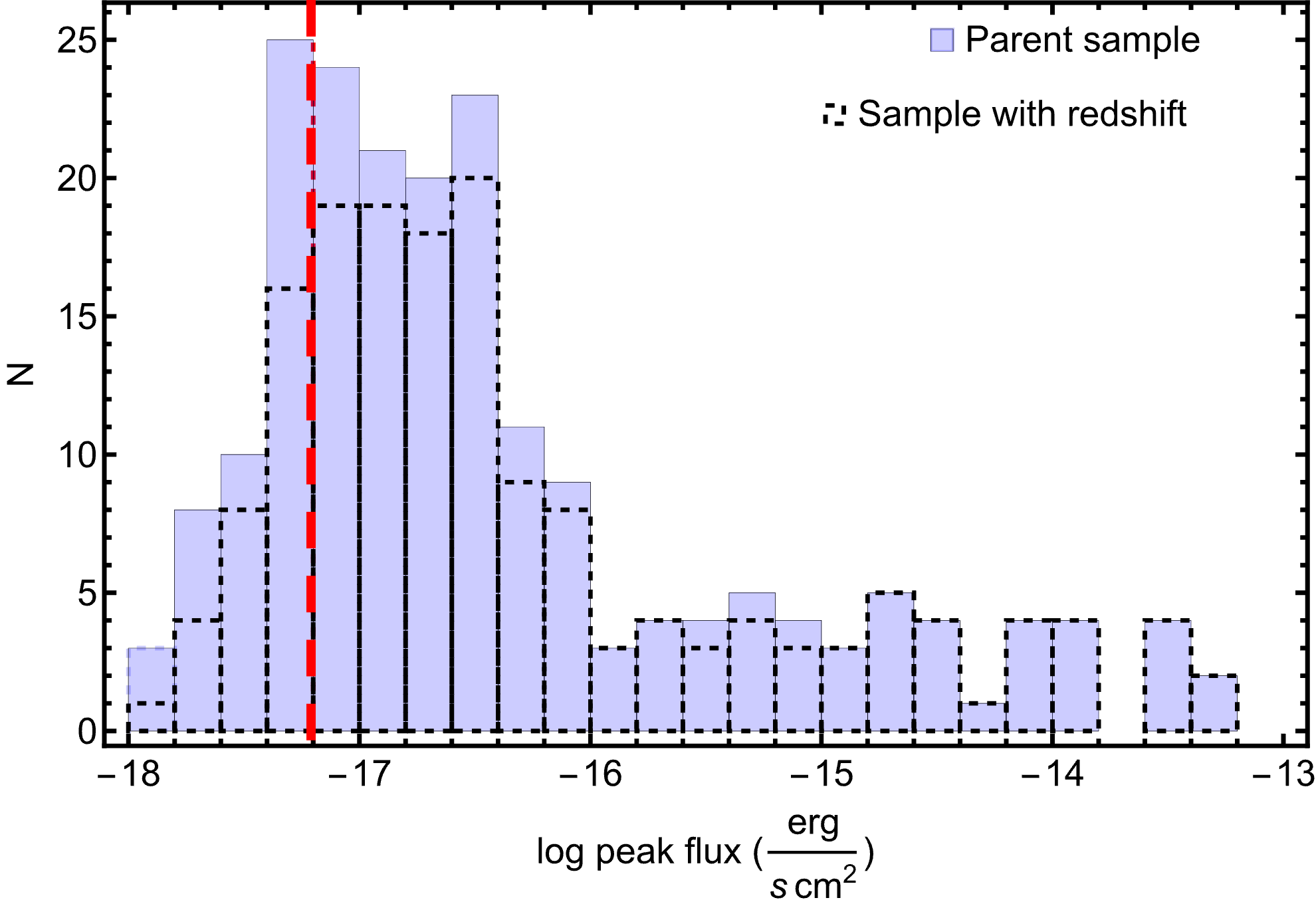} & \includegraphics[width = 0.35\textwidth]{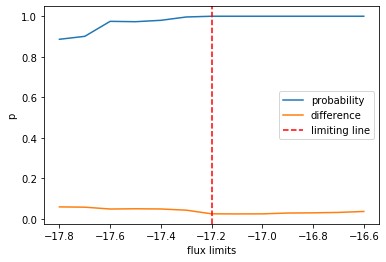} \\  
    \end{tabular}
    \caption{(\textbf{Left}): peak flux distribution for ``parent'' sample and subsample with known redshift. Limiting flux shown in red. (\textbf{Right}): plot of probability (blue) and distance between samples as given by the KS test (orange) as a function of flux limit. Limiting line $f_{\rm lim}$ shown in red.}
    \label{fig:fluxhist}
\end{figure}

We show the distribution of the parent sample and subsample in the left panel of Figure \ref{fig:fluxhist}, with the limiting line shown in red. We plot this probability as a function of flux limit (blue), as well as the distance between the distributions (orange), in the right panel of Figure \ref{fig:fluxhist}. 

\subsection{Removing Selection Bias and Redshift Evolution}\label{bias}

After defining the limiting lines, we can then mimic the evolution of this variable with redshift using a simpler function of redshift, ${f(z)}=\frac{1}{(1+z)^\delta}$. We here adopt the choice of a simple function, but in principle it is possible to use a more complex function as already shown in~\citep{2015MNRAS.451.3898D} and obtain compatible results. Using a modified version of Kendall's $\tau$ statistics, we can compute an evolutionary function $f(z)$ where the slope of the function is defined by $\delta$. The values for $\delta$ where $\tau = 0$ represent the removal of evolutionary effects. We define errors on $\delta$ out to 1 $\sigma$, corresponding to the $\delta$ values where $\tau=\pm 1$.

%%%%%%%%%%%%%%%%%%%%%%%%%%%%%%%%%%%%%%%%%
\section{Results}\label{results}

We here clarify that the purpose of our analysis is to show how similar our results are, compared to other ones in the literature given that our sample size differs from other studies for $E_{\rm iso}$ and $T^{*}_{90}$ (our sample has 80 GRBs), while this is the first time in the literature that we compare the results for the $L_{\rm a,radio}$ and \rm $T_{\rm a, radio}^*$ (our sample has 18 GRBs) with the previous results in the literature performed in X-rays and optical. This is an essential comparison to allow the determination of the intrinsic nature of the $L_{\rm a, radio}$ and $T^{*}_{\rm a, radio}$ correlation and to check for the universality of the results related to the evolutionary functions for these variables with the EP method.

\begin{figure}[H]
    \begin{tabular}{cc}
        \includegraphics[width = 0.35\textwidth]{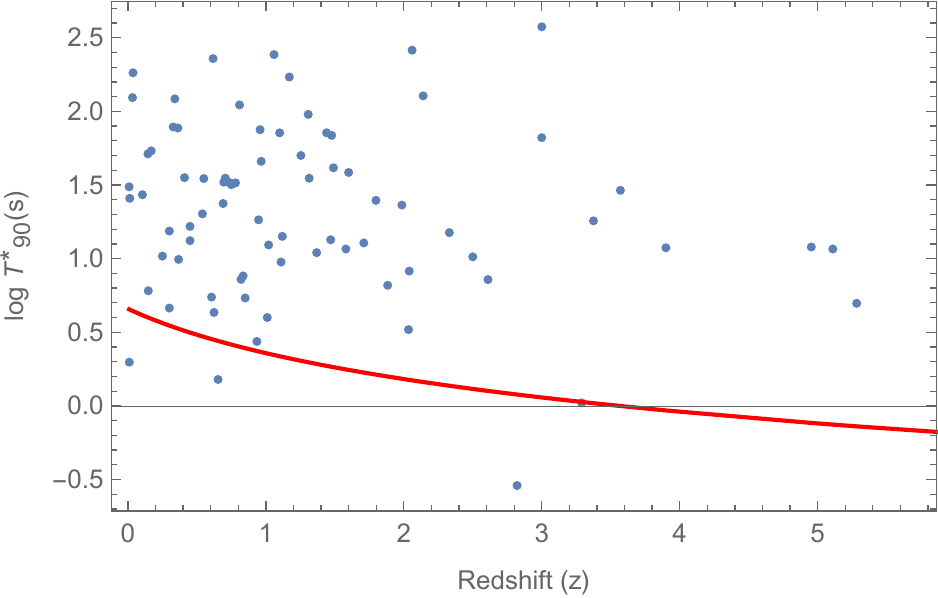} & \includegraphics[width = 0.35\textwidth]{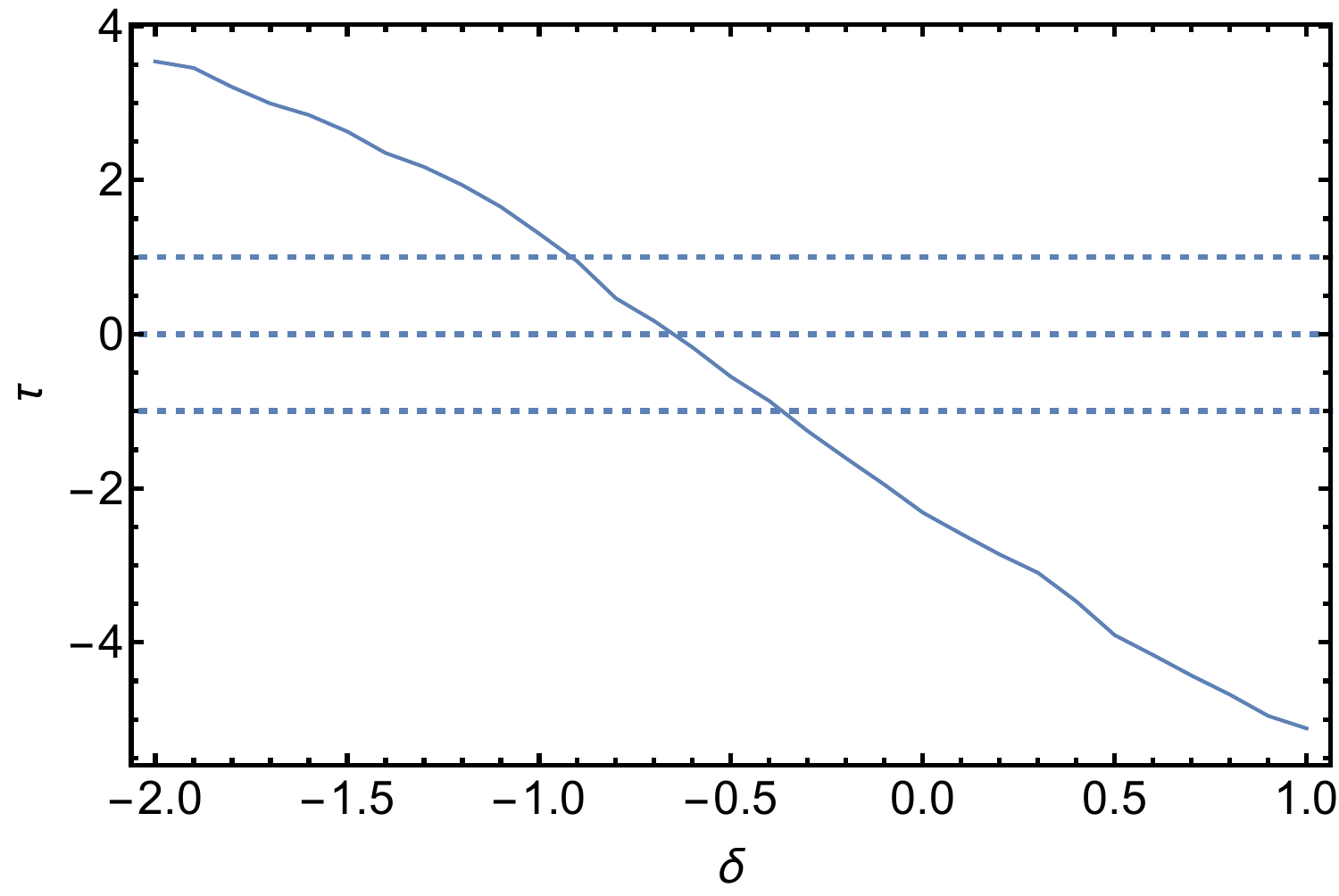} \\ 
        \includegraphics[width = 0.35\textwidth]{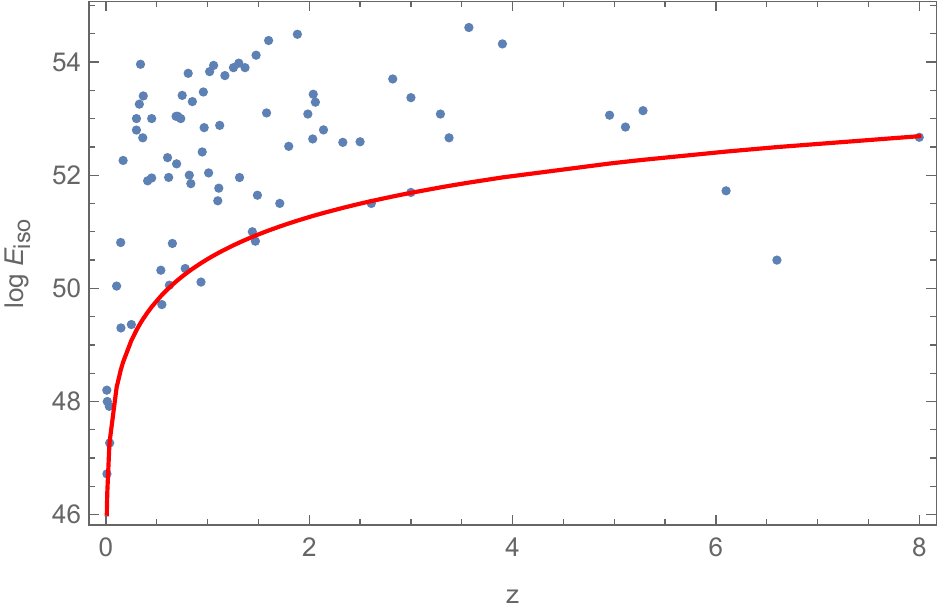} & \includegraphics[width = 0.35\textwidth]{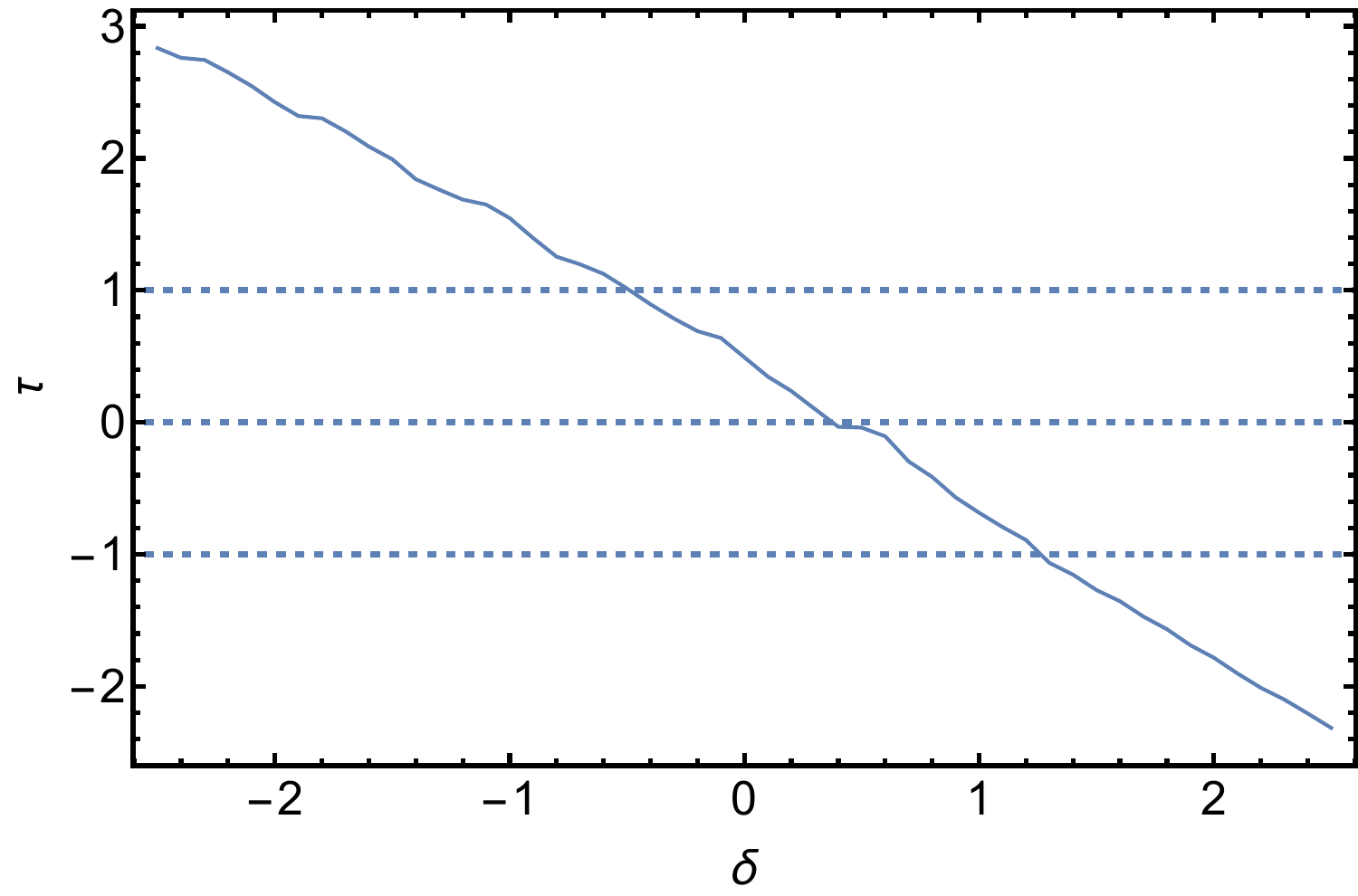}
    \end{tabular}
    \caption{{The} %hyphen in figures should be minus, please confirm whether it is necessary to replace the pictures
    %DL: unclear which hyphen you are referring to?
(\textbf{upper left}) panel shows the values of $\log T_{90}^*$ vs. redshift in blue and the limiting $\log T_{90}^*$ in the rest-frame in red. The (\textbf{upper right}) panel shows the Kendall $\tau$ vs. the slope of the evolutionary function with 1 $\sigma$ errors shown with dashed blue lines. As with the upper panels, the (\textbf{lower left}) panel shows values of $\log E_{\rm iso}$ vs. redshift in blue with the limiting line in red, and the (\textbf{right}) panel shows the Kendall $\tau$ vs. slope of the evolutionary function with 1 $\sigma$ errors as dashed blue lines.}
    \label{fig:EisoT90}
\end{figure}

Using the procedure outlined above, we correct $\log E_{\rm iso}$, $\log T_{90}^*$ using the formulation $\log E_{\rm iso}^{'} = \log E_{\rm iso}- \log( (1+z)^{\delta_{E_{\rm iso}}})$ and $\log T_{90}^{*'} = \log T_{90}^* - \log((1+z)^{\delta_{T_{90}^*}})$, where all quantities that have $'$ symbol are the variables for which the evolution has been removed, thus they are no longer dependent on the redshift. We find $\delta_{T_{90}^*} = -0.65 \pm 0.27$, with the 1 $\sigma$ errors defined as the average of the values of $\tau = 1$ and $\tau = -1$, and $\delta_{E_{\rm iso}} = 0.39 \pm 0.88$. Figure \ref{fig:EisoT90} shows these results---the top left panel shows $\log T_{90}^*$ for the sample of 80 GRBs as a function of redshift, with the limiting value shown in red. The top right panel highlights the evolutionary function for $\log T_{90}^*$, with dashed lines at $\tau=0, \pm 1$. The same plots for $\log E_{\rm iso}$ are shown in the bottom panels.

\begin{figure}[H]
    \begin{tabular}{cc}
        \includegraphics[width = 0.35\textwidth]{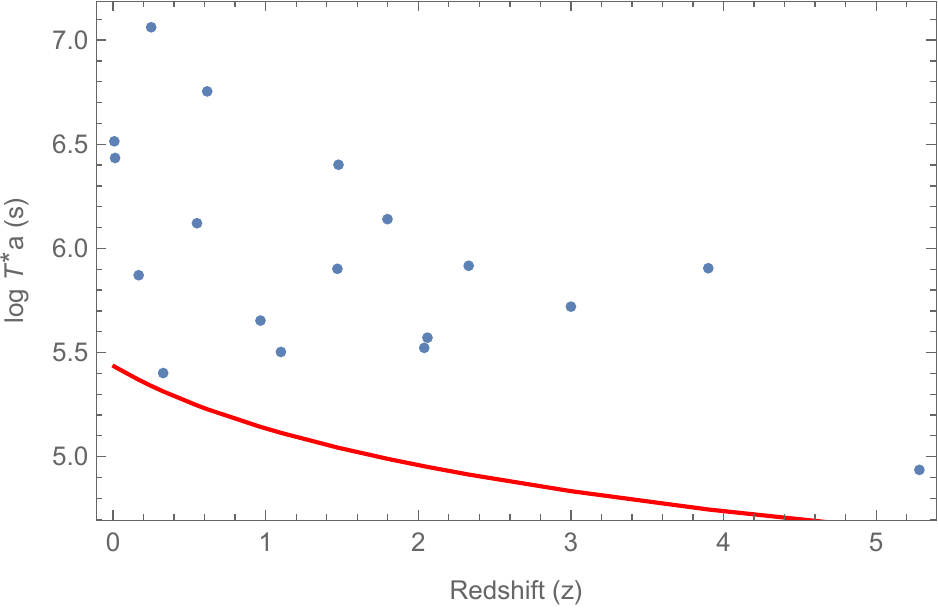} & \includegraphics[width = 0.35\textwidth]{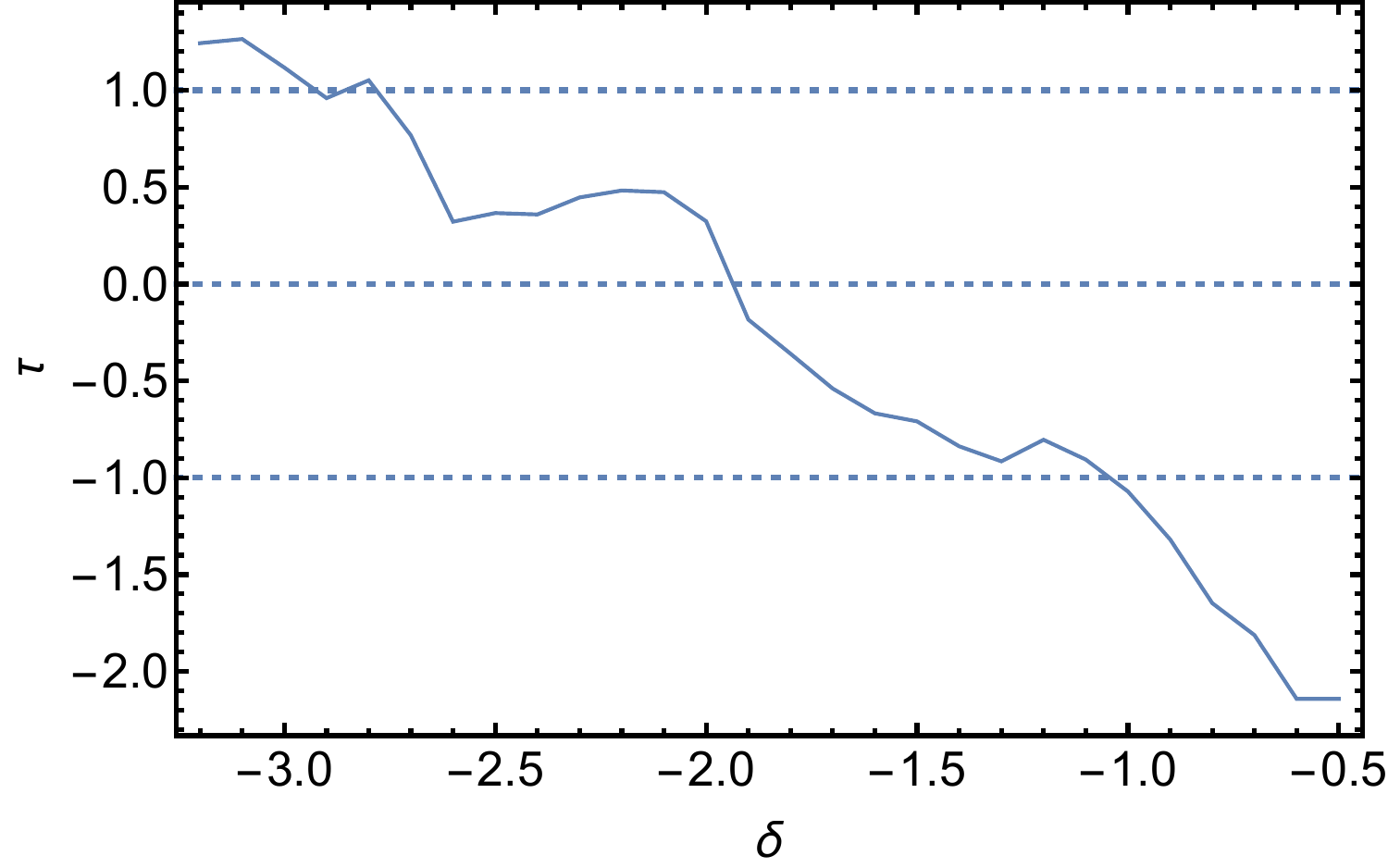} \\ \includegraphics[width = 0.35\textwidth]{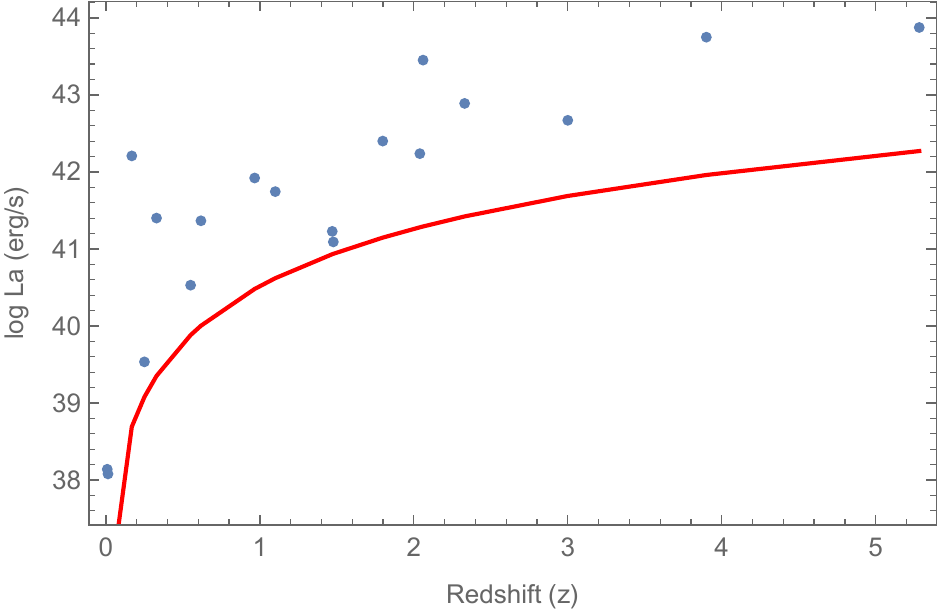} & \includegraphics[width = 0.35\textwidth]{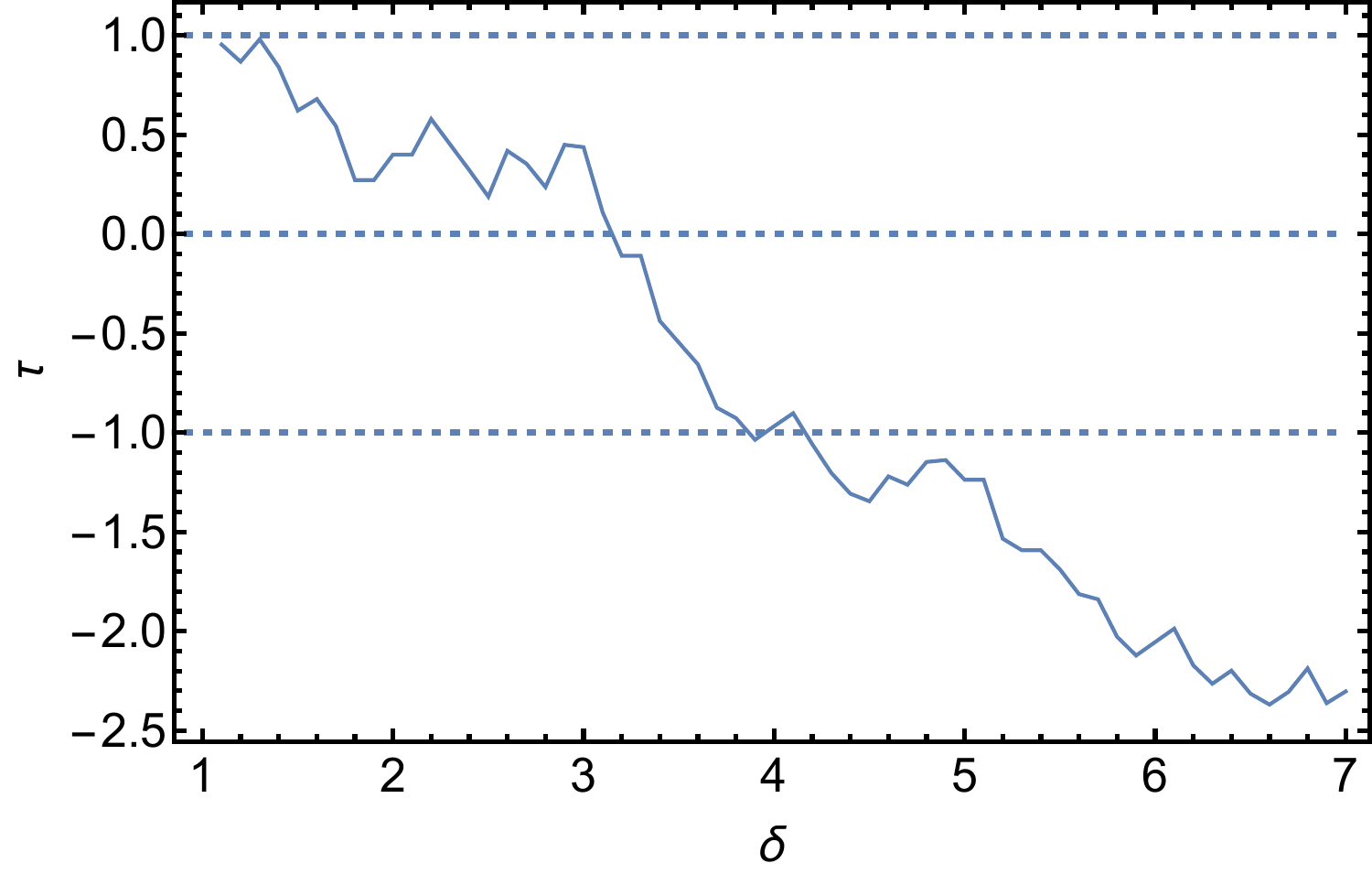}
    \end{tabular}
    \caption{The (\textbf{upper left}) panel shows the values of $\log T_{\rm a, radio}^*$ vs. redshift in blue and the limiting $\log T_{\rm a, radio}^*$ in the rest-frame in red. The (\textbf{upper right}) panel shows the Kendall $\tau$ vs. the slope of the evolutionary function with 1 $\sigma$ errors shown with dashed blue lines. As with the upper panels, the (\textbf{lower left}) panel shows values of $\log L_{\rm a, radio}$ vs. redshift in blue with the limiting line in red, and the (\textbf{right}) panel shows the Kendall $\tau$ vs. slope of the evolutionary function with 1 $\sigma$ errors as dashed blue lines.}
    \label{fig:LaTa}
\end{figure}

For the plateau sample of 18 GRBs, we use the same formulation to obtain $\log L_{\rm a, radio}^{'} = \log L_{\rm a, radio} - \log((1+z)^{\delta_{L_{\rm a, radio}}})$ and $\log T_{\rm a, radio}^{*'} =  \log T_{\rm a, radio}^* - \log((1+z)^{\delta_{T_{\rm a, radio}^*}})$. We find the values of $\delta$ for $\log L_{\rm a, radio}$ and $\log T_{\rm a, radio}^*$ as $\delta_{T_{\rm a, radio}^*} = -1.94 \pm 0.86$ and $\delta_{L_{\rm a, radio}} = 3.15 \pm 1.65$. These results are shown in Figure \ref{fig:LaTa}---$\log T_{\rm a, radio}^*$ as a function of redshift is shown in the top left panel, with the limiting values in red. The evolutionary function is shown in the top right panel, with dashed blue lines at $\tau = 0, \pm 1$. The bottom two panels display the same plots for $\log L_{\rm a, radio}$. 

%%%%%%%%%%%%%%%%%%%%%%%%%%%%%%%%%%%%%%%%%%
\section{Discussion and Conclusions}\label{discussion}

Using a sample of 80 GRBs with observed radio afterglow, we test the use of the EP method on $\log E_{\rm iso}$ and $\log T^{*}_{90}$ for a subsample 80 GRBs, and on $\log L_{\rm a, radio}$ and $\log T_{\rm a, radio}^*$ for a subsample of 18 GRBs. We find that when considering $\log E_{\rm iso}$ and $\log T^{*}_{90}$, we obtain indices for the parameter of redshift evolution, $\delta$, of $\delta_{T_{90}^*} = -0.65 \pm 0.27$ and $\delta_{E_{\rm iso}} = 0.39 \pm 0.88$, while the values of $\delta$ for $\log L_{\rm a, radio}$ and $\log T_{\rm a, radio}^*$ are $\log T_{\rm a, radio}^*$ as $\delta_{T_{\rm a, radio}^*} = -1.94 \pm 0.86$ and $\delta_{L_{\rm a, radio}} = 3.15 \pm 1.65$.

For $\log T_{90}^*$, for $\log T_{\rm a, radio}$, and $\log L_{\rm a, radio}$, we find relatively strong evolution of each variable with redshift. The luminosity presents the strongest correlation with redshift, emphasizing the necessity of correction for these effects before using data in correlation analysis. $E_{\rm iso}$, by contrast, appears to be the most independent from redshift, with the smallest value for $|\delta|$. 

We find that our values are comparable to values obtained in previous studies. A study by~\citet{2019MNRAS.488.5823L} of the cosmological evolution of isotropic energy $E_{\rm iso}$, burst duration $T_{90}^*$, jet opening angle $\theta_j$, and luminosity $L_j$ reports $\delta$ values compatible with our findings $T_{90}^*$ and  $L_{\rm a, radio}$ within 1 $\sigma$, and values compatible with our $\delta_{E_{\rm iso}}$ within approximately 2 $\sigma$. Specifically, they find a value of $\delta_{E_{\rm iso}} = 2.3 \pm 0.5$, which agrees with our value of $\delta_{E_{\rm iso}} = 0.39 \pm 0.88$ within 2.17 $\sigma$. We also see agreement with $\delta_{T_{90}^*} = -0.8 \pm 0.3$, with a 0.55 $\sigma$ difference from our value of $\delta_{T_{90}^*} = -0.65 \pm 0.27$, and in the luminosity with  $\delta_{L_{\rm j}} = 3.5 \pm 0.5$, a 0.2 $\sigma$ difference from our value of  $\delta_{L_{\rm a, radio}} = 3.15 \pm 1.65$. 

These results also agree with previous values of $\delta$ for $L_{\rm a}$ and $T_{\rm a}^*$ in X-ray and optical wavelengths.~\citet{2013ApJ...774..157D} conduct a similar analysis of the luminosity and rest-frame end time of the plateau emission using X-ray data. Their value for correction for $\delta_{T_{\rm a}^*}$, reported as $\delta_{T_{\rm a, X}^*} = -0.85 \pm 0.3$, is compatible with our value of $\delta_{T_{\rm a, radio}^*} = -1.94 \pm 0.86$ within 1.23 $\sigma$. However, they find a very slow evolution in luminosity, with a value of $\delta_{L_{\rm a, X}} = -0.05 \pm 0.35$, which is a 1.94 $\sigma$ difference from our value of $\delta_{L_{\rm a, radio}} = 3.15 \pm 1.65$. This discrepancy is likely due in part to the small sample size, which may exaggerate the extent of the evolution present in our sample, but may also be due to differences in the behavior of the X-ray and radio emission.

We have corrected the luminosity and time in X-rays with 222 GRB lightcurves with a given redshift, presenting plateaus according the~\citet{2007ApJ...662.1093W} model and in optical with 181 GRBs with plateaus taken from~\citet{2020ApJ...905L..26D}, but with the additional analysis of 80 GRBs found in the literature. For tackling this analysis, we followed the same procedure described in the current paper. The results of this analysis reports $\sigma$, and $\delta_{L_{\rm a, X}} = 2.42 \pm 0.58$, which is a 0.44 $\sigma$ difference from our value. They also report values in optical of $\delta_{T_{\rm a, opt}^*} = -2.11 \pm 0.49$ and $\delta_{L_{\rm a, opt}} = 3.96 \pm 0.58$, which both agree with our result within 1 $\sigma$. 

In general, it can be seen that a larger sample size is preferred when applying the EP method. In our case, for the sample pertinent to $E_{\rm iso}$, and $T_{90}^*$ we choose the limiting values while excluding $\leq 10\%$ of the overall sample. However, for the smaller sample of 18 plateau GRBs, the limiting values are chosen so that we do not exclude any data points due to the small sample size. In addition, this conservative choice would allow us to have smaller error bars on the slope of the evolutionary functions. However, the $\delta$ values obtained for $L_{\rm a, radio}$ and $T_{\rm a, radio}^*$ are similar to values found in previous studies of larger sample sizes, thus indicating that the EP method is still successful even with a small~dataset. 

GRB correlations in radio afterglows related to the plateau emission are crucial to understand if the jet break is coincident with the end of the plateau emission. To investigate this point, a multiwavelength analysis not only in optical and X-rays must be performed together with the radio data. Since the evolution of the variables with redshift can change the time at which the break happens, it is crucially important to correct for the redshift evolution.
This study is also the preliminary step to the investigation of the intrinsic nature of the plateau emission correlations.

After our analysis on the radio observations both for the prompt emission in relation to the variables of $E_{\rm iso}$ and $T^{*}_{90}$ and for the afterglow emission in relation to $L_{\rm a,radio}$, $T^{*}_{\rm a,radio}$,
we can conclude: 
\begin{enumerate}
    \item {After testing} intrinsic properties of a GRB, such as $E_{\rm iso}$ and $T_{90}^*$, as well as properties such as $L_{\rm a, radio}$ and $T_{\rm a, radio}^*$, we see $T_{90}^*$, $L_{\rm a, radio}$ and $T_{\rm a, radio}^*$ present strong correlation with redshift, thus indicating that they are susceptible to redshift evolution. %Is the bold necessary? please confirm or define. %DL: no, it can be removed
    \item The $\delta$ values obtained in this work agree with those of previous studies, indicating that this trend of strong correlation with redshift still holds true in radio wavelengths.
    \item The study of these evolutionary functions is the first necessary step to determine the true intrinsic nature of the correlations, object of our study.
\end{enumerate}

\vspace{6pt}
\authorcontributions{%
Conceptualization: M.D.; Methodology: M.D.; Software: M.D.; Validation: M.D. and D.L.; Formal Analysis: D.L.; Investigation, D.L. and M.D.; Data Curation: P.C.; Writing---Original Draft Preparation: D.L.; Writing---Review and Editing Draft: M.D., N.F. and D.L.; Supervision: M.D. and N.F. All authors have read and agreed to the published version of the manuscript.}

\funding{{This research was funded by the National Astronomical Observatory of Japan (NAOJ).}}%Please add: “This research received no external funding” or “This research was funded by NAME OF FUNDER grant number XXX.” and  “The APC was funded by XXX”. Check carefully that the details given are accurate and use the standard spelling of funding agency names at \url{https://search.crossref.org/funding}, any errors may affect your future funding.}
%MDPI: Please check the accuracy of funding data and any other information carefully.

\institutionalreview{{Not applicable.}}%In this section, please add the Institutional Review Board Statement and approval number for studies involving humans or animals. Please note that the Editorial Office might ask you for further information. Please add “The study was conducted according to the guidelines of the Declaration of Helsinki, and approved by the Institutional Review Board (or Ethics Committee) of NAME OF INSTITUTE (protocol code XXX and date of approval).” OR “Ethical review and approval were waived for this study, due to REASON (please provide a detailed justification).” OR “Not applicable” for studies not involving humans or animals. You might also choose to exclude this statement if the study did not involve humans or animals.}

\informedconsent{{Not applicable. }}%Any research article describing a study involving humans should contain this statement. Please add “Informed consent was obtained from all subjects involved in the study.” OR “Patient consent was waived due to REASON (please provide a detailed justification).” OR “Not applicable” for studies not involving humans. You might also choose to exclude this statement if the study did not involve humans.

%Written informed consent for publication must be obtained from participating patients who can be identified (including by the patients themselves). Please state “Written informed consent has been obtained from the patient(s) to publish this paper” if applicable.}

\dataavailability{The data are taken from~\citet{2012ApJ...746..156C, 2013ApJ...767..161Z, 2015ApJ...814....1L, 2019ApJ...884..121L, 2015ApJ...812..122C, 2019MNRAS.486.2721B, 2014MNRAS.440.2059A, 2015ApJ...806...52S, 2018ApJ...859..134L, 2020ApJ...894...43K, 2016ApJ...833...88L, 2020ApJ...891L..15C, 2017ApJ...845..152B, 2017Sci...358.1579H, 2018Natur.554..207M, 2018ApJ...867...57R, 2018ApJ...856L..18M, 2021ApJ...907...60M} and \mbox{\citet{2020MNRAS.496.3326R}}.}

\acknowledgments{%
This work was made possible in part by the United States Department of Energy, Office of Science, Office of Workforce Development for Teachers and Scientists (WDTS) under the Science Undergraduate Laboratory Internships (SULI) program. We thank Cuellar for managing the SULI program at Stanford National Accelerator Laboratory. We also acknowledge the National Astronomical Observatory of Japan for their support in making this research possible, as well as Kevin J. Zvonarek for his help in preparing the dataset in this work. NF acknowledges financial  support  from UNAM-DGAPA-PAPIIT  through grant IN106521. D. Levine acknowledges support from NAOJ division of Science.}

\conflictsofinterest{%
The authors declare no conflict of interest. The founding sponsors had no role in the design of the study; in the collection, analyses, or interpretation of data; in the writing of the manuscript, and in the decision to publish the results.}

\end{paracol}
\reftitle{References}

\end{document}